\documentclass[aps,prb,twocolumn,showpacs,amsmath,amssymb]{revtex4}
\usepackage{graphicx}
\usepackage{dcolumn}
\usepackage{bm}

\begin{document}
\title{Electronic Structure and Exchange Interactions of Na$_{2}$V$_{3}$O$_{7}$}
\author{V.V. Mazurenko$^{1,2,3}$, F. Mila$^{2}$ and V.I. Anisimov$^{1,3}$}
\affiliation{$^{1}$Theoretical Physics and Applied Mathematics Department, Urals State 
Technical University, Mira Street 19,  620002
Ekaterinburg, Russia \\
$^{2}$Institute of Theoretical Physics, Ecole Polytechnique F\'ed\'erale de Lausanne, CH-1015 
Lausanne, Switzerland \\
$^{3}$Institute of Metal Physics, Russian Academy of Sciences, 620219 Ekaterinburg GSP-170, Russia}
\date{\today}

\begin{abstract}
We have performed first-principle calculations of the electronic structure and exchange couplings 
for the nanotube compound 
Na$_{2}$V$_{3}$O$_{7}$ using the LDA+U approach. Our results show that while the intra-ring exchange 
interactions are mainly antiferromagnetic,  the inter-ring couplings are {\it ferromagnetic}. We argue that
this is a consequence of the strong hybridization between  filled and vacant 3d vanadium orbitals 
due to the low symmetry of  Na$_{2}$V$_{3}$O$_{7}$, which results 
into strong - and often dominant - ferromagnetic contributions to the total exchange interaction
between vanadium atoms. A comparison with results of previous works is included.
\end{abstract}

\maketitle
\section{Introduction}
The values of the  exchange interactions between the magnetic moments carried by
metallic atoms constitute a crucial piece of information  
for the theoretical investigation of the magnetic properties 
(magnetization, susceptibility, specific heat, etc.) of low-dimensional 
quantum S=$\frac{1}{2}$ systems, for instance vanadates.\cite{Frederic1,Frederic2,Korotin99} 
There are several ways to obtain this information. 
One of them is based on the estimation of the 
exchange integrals from the hopping parameters obtained by fitting the band-structure
with an extended tight-binding 
model \cite{moskvin} or  through a downfolding \cite{Roser} and projection \cite{AnisimovWF} 
procedure.
The first-principle method for the calculation of the exchange interaction parameters 
was proposed by Lichtenstein {\it{et al.}} \cite{Liechtenstein}
Within this scheme the exchange interaction parameters are determined 
by calculating the second variation of the
total energy  $\delta^2 E$ with respect to 
small deviations of the magnetic moments from the collinear magnetic configuration. 

The first theoretical investigations \cite{whangbo,saha-dasgupta} 
of the magnetic couplings of Na$_{2}$V$_{3}$O$_{7}$, a Mott insulator with the
rather exotic topology of  a nanotube, were based on the former method. They relied on different
estimates of the hopping integrals and lead to conflicting results. 
Whangbo and Koo \cite{whangbo} have employed a spin dimer analysis in order to estimate 
the relative magnitudes of the spin  exchange interaction parameters expected for Na$_{2}$V$_{3}$O$_{7}$. They have concluded that Na$_{2}$V$_{3}$O$_{7}$ could be described by six helical spin chains 
and should show a gap in the spin excitations.

Dasgupta {\it{et al.}} \cite{saha-dasgupta} have performed the first {\it ab-initio} microscopic analysis 
of the electronic and magnetic properties of 
Na$_{2}$V$_{3}$O$_{7}$ using linear augmented plane wave (LAPW) method.\cite{LAPW}  
Since, due to the geometry, the 3d {\it xy} bands of the vanadium atoms 
are well separated from other states, these authors have derived a V 3d {\it xy} orbital model 
Hamiltonian using the downfolding procedure within the 
framework of the N-th order muffin-tin
orbital method.\cite{Andersen2000} Based on the calculated hopping parameters
between {\it {xy}} orbitals, they have estimated the exchange interactions between vanadium 
atoms in Na$_{2}$V$_{3}$O$_{7}$ using standard superexchange theory and 
proposed an effective Heisenberg model. The calculated magnetic susceptibility as a 
function of temperature determined by exact diagonalization of finite clusters 
is in good agreement  with experimental data.

In this paper we report a detailed investigation of the isotropic 
exchange interactions in Na$_{2}$V$_{3}$O$_{7}$ based on 
a first-principle LDA+U calculation of its electronic and magnetic structure. 
The exchange interactions between vanadium 
atoms were calculated using the Green-function method developed by 
Lichtenstein {\it{et al.}} \cite{Lichten} In order to provide a microscopic explanation 
of the resulting exchange interactions, which are {\it qualitatively} different from previous 
estimates, we have used an extended Kugel-Khomskii model 
that includes the hybridization between filled and vacant orbitals
using the hopping parameters calculated by the projection procedure.
We conclude that estimating the exchange interactions of a system with a very exotic
geometry such as Na$_{2}$V$_{3}$O$_{7}$ within simple superexchange
between {\it xy} orbitals with on-site Coulomb interaction is not possible.

The paper is organized as follows. In Section II we shortly describe
the crystal structure of Na$_{2}$V$_{3}$O$_{7}$ and present the results of the LDA calculation. 
In section III, we present the results of the LDA+U calculation,  and we discuss
the origin of the exchange interaction parameters. 
In Section IV, we briefly summarize our results.

\section{LDA calculation}
A simplified view of the crystal structure of Na$_{2}$V$_{3}$O$_{7}$ is presented in Fig.1.
The main building blocks of the crystal structure are distorted vanadium-oxygen pyramids. 
Vanadium atoms are roughly in the center of a pyramid
which is built by five oxygen atoms. There are three types of vanadium atoms V1-V2-V3 which 
form a basic structure unit.
Three of these units connected by edges are needed to form a ring of the tube.
The lowest energy orbital is the V 3d orbital of $xy$ symmetry \cite{Korotin99} (using a convention 
where the axes of the coordinate system are directed towards the oxygen ions in the plane), 
which is the orbital whose lobes point away from the oxygen. More detailed information 
about the crystal structure of Na$_{2}$V$_{3}$O$_{7}$ can be found in Ref.\onlinecite{millet}.
\begin{figure}[h]
\centering
\includegraphics[width=0.5\textwidth]{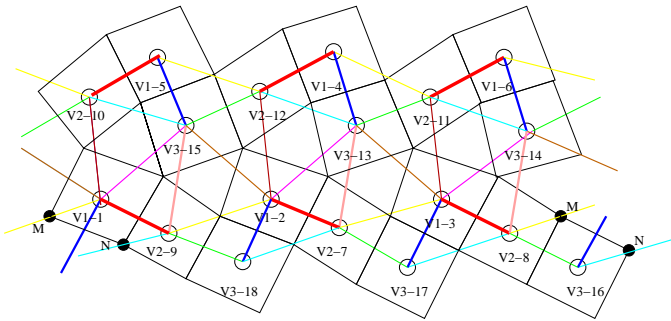}
\caption {(Color online) Simplified crystal structure and interaction paths between 
vanadium atoms in Na$_{2}$V$_{3}$O$_{7}$.
Equivalent interaction paths have the same color. The tube structure is made by putting 
M and N into contact.}
\label{total}
\end{figure}

The electronic structure calculation of Na$_{2}$V$_{3}$O$_{7}$ was performed using the 
Tight Binding Linear-Muffin-Tin-Orbital 
Atomic Sphere Approximation (TB-LMTO-ASA) method in terms of the conventional 
local-density approximation. \cite{Andersen} 
The band structure of Na$_{2}$V$_{3}$O$_{7}$ obtained from LDA calculations is presented in Fig.2.
There are 18 bands near the Fermi level which have $xy$ symmetry 
(since there are 18 vanadium atoms in the unit cell) and 
are separated from the other V 3d bands as well as from the 2p oxygen
bands. The corresponding partial densities of states are shown in Fig.3.
\begin{figure}
\centering
\includegraphics[angle=270,width=0.45\textwidth]{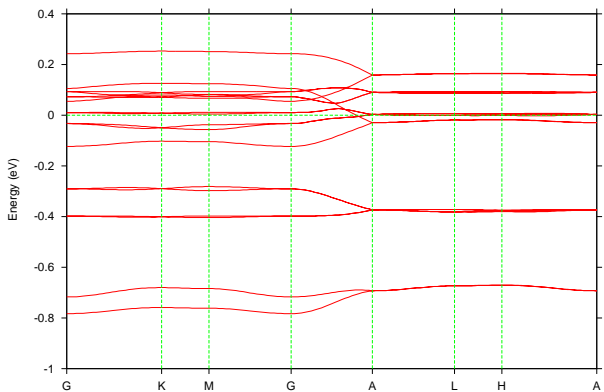}
\caption {Band structure of Na$_{2}$V$_{3}$O$_{7}$ near the Fermi level (0 eV).}
\end{figure}
Six bands have an energy significantly lower than the others, which are located near the Fermi level. 
Two of them have an energy near -0.7 eV, while the other four are located between -0.3 and -0.4 eV.
Such a band location of active orbitals (occupied by electrons) is unusual in comparison
 with the band structure of other S=$\frac{1}{2}$ transition metal compounds. \cite{tellurid}
These bands are in good agreement with those presented in Ref.~\onlinecite{saha-dasgupta}.

With one subset of energy bands well separated from the rest, one can expect that a tight-binding 
model with a single
{\it xy}-orbital per V site (i.e. a 18-band model) 
will provide a good
approximation to the full band structure close to the Fermi level.
\vspace{1cm}
\begin{figure}[h]
\centering
\includegraphics[angle=0,width=0.45\textwidth]{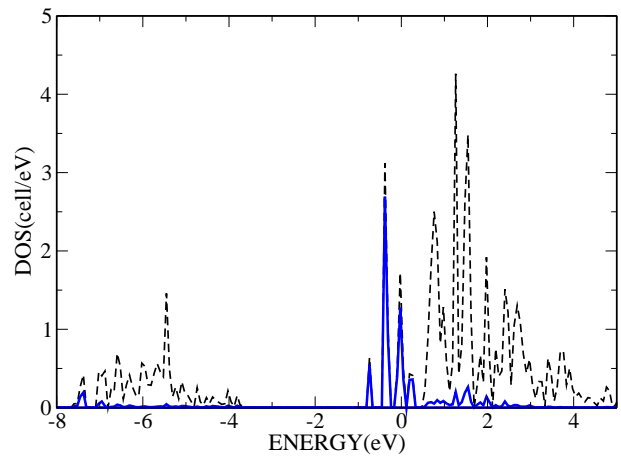}
\caption {(Color online) Partial density of 3d vanadium states obtained from LDA calculations.
The solid blue and the dashed black lines are the density of states of 3d$_{xy}$ and the 
total density of 3d states, respectively.}
\end{figure}
For this we express the LDA band structure results in terms of the low-energy model Hamiltonian
\begin{eqnarray}
H_{TB} = \sum_{i, j, \sigma} t_{ij} a^{+}_{i \sigma} a_{j \sigma} ,
\end{eqnarray}
where {\it i} and {\it j} are site indexes, $t_{ij}$ is the  effective hopping integral between sites $i$ and $j$,
and $a^{+}_{i \sigma}$ is the operator which creates an electron with a spin $\sigma$
in a state with a Wannier function centered at the {\it i}th atom. In order to obtain the hopping 
parameters $t_{ij}$, 
we have employed the projection procedure. \cite{AnisimovWF} According to this procedure, the
hopping integrals in the Wannier function basis
$|W^{m}_{i}\rangle$ ({\it {m}} is orbital index, {\it {i}} is site index), which is defined as the Fourier transform
of a certain linear combination of Bloch functions $|\psi_{nk}\rangle$ ($n$ is the band index and k is 
the wave-vector in reciprocal space) 
can be written in the general form:
\begin{eqnarray}
\label{E_WF}
t^{mm'}_{ij} & = & \langle W^{m}_{i}|\biggl(\sum_{n, k}|\psi_{nk}\rangle \epsilon_{n}(k)  
\langle \psi_{nk}|\biggr)|W^{m'}_{j}\rangle  \nonumber \\
\end{eqnarray}
where $\epsilon_{n}(k)$ is the eigenvalue in the basis of linearized muffin-tin orbitals.
The calculated inter-ring and intra-ring hopping parameters between 3d {\it xy} orbitals of vanadium 
atoms are presented in Table I.
\begin {table}
\centering
\caption [Bset] {Calculated hopping integrals between {\it xy} orbitals of vanadium atoms for 
the one-orbital model and estimated 
exchange interaction parameters $J^{xy}_{ij}$ ($J^{xy}_{ij}=\frac{4 \, (t^{xy}_{ij})^{2}}{U}$, $U = 2.3\ eV$) 
of Na$_{2}$V$_{3}$O$_{7}$ (in meV). }
\label {basisset}
\begin {tabular}{lccccc}
  \hline
  i-j  & \quad \quad \quad t$^{xy}_{ij}$  & \quad \quad \quad J$^{xy}_{ij}$   \\
  \hline
  (V1-1)-(V2-9)  & \quad \quad \quad -169  & \quad \quad \quad  49.67          \\
  (V1-1)-(V3-16) & \quad \quad \quad -145  & \quad \quad \quad 36.56          \\
  (V1-1)-(V2-8)  & \quad \quad \quad -135  & \quad \quad \quad 31.70          \\
  (V2-7)-(V3-18) & \quad \quad \quad -138  & \quad \quad \quad 33.12          \\
  (V2-9)-(V3-18) & \quad \quad \quad -118  & \quad \quad \quad 24.22          \\
  (V2-9)-(V3-15) & \quad \quad \quad -36   & \quad \quad \quad 2.25           \\
  (V1-1)-(V2-10) & \quad \quad \quad -32   & \quad \quad \quad 1.78           \\
  (V1-1)-(V3-15) & \quad \quad \quad  -6   & \quad \quad \quad 0.06           \\
  (V1-2)-(V3-15) & \quad \quad \quad -15   & \quad \quad \quad 0.39           \\
  \hline
\hline
\end {tabular}
\end {table}
These values are in excellent agreement with the results of the downfolding procedure. \cite{saha-dasgupta}
One can see that intra-ring hopping integrals are in the range -120 meV to -170 meV.
The inter-ring hopping integrals between occupied {\it xy} orbitals of vanadium atoms are smaller
than the intra-ring ones. 

Having set up the picture of the dynamics of individual electrons we shall now introduce 
their on-site Coulomb interaction $U$ to obtain the one-band Hubbard model
\begin{eqnarray}
H_{Hub} = \sum_{i, j, \sigma} t_{ij} a^{+}_{i \sigma} a_{j \sigma} + \frac{U}{2} \sum_{i, \sigma}
 n_{i \sigma} n_{i -\sigma} ,
\end{eqnarray}
where $n_{i \sigma} = a^{+}_{i \sigma} a_{i \sigma}$ is the particle-number operator.
Based on the obtained hopping integrals, we have estimated the exchange interaction parameters using 
the expression $J^{xy}_{ij}=\frac{4 \, (t^{xy}_{ij})^{2}}{U}$ (see Table I), which appears upon 
mapping the one-band Hubbard model onto the Heisenberg model in 
the limit $U>>t$, and which only describes the antiferromagnetic superexchange process due to the 
hybridization between non-degenerate orbitals. 
The sum in Eq.(2) runs over all bands. Therefore in the {\it xy}-orbital model we take into account 
all interaction paths between $xy$ orbitals on different vanadium atoms. 

In order to check the reliability of the mapping of the band structure onto the 
one orbital tight-binding model (Eq.1), we have also derived a five-orbital model described by a 
tight-binding Hamiltonian that includes five orbitals per V site:
\begin{eqnarray}
H^{full}_{TB} = \sum_{\substack {i, j, \sigma \\ m, m'}} t^{mm'}_{ij} a^{+}_{i m \sigma} a_{j m' \sigma} 
\end{eqnarray}
using the projection procedure to calculate the hopping integrals. The results of this
 calculation are presented in Table II.
\begin {table}[h]
\centering
\caption [Bset] {Calculated hopping integrals between the  $3d$ orbitals of vanadium atoms including
five orbitals per V site model (in eV). 
$t_{ij}^{mm}$ corresponds to the hopping process between the filled {\it xy} orbitals of {\it i}th 
and {\it j}th vanadium atoms, while
$t_{ij}^{mm'}$corresponds to the hopping parameters between filled ($m$) and vacant ($m'$) orbitals.}
\label {basisset}
\begin {tabular}{lccccc}
  \hline
  i-j  & $t^{mm}_{ij}$  & $I^{a}_{ij} =(t^{mm}_{ij})^{2}$   & \quad $I^{f}_{ij}=\sum_{\substack {m  m' \\ (m \neq m')}} (t^{mm'}_{ij})^{2}$ \\
  \hline
  (V1-1)-(V2-9)  & -0.166  & 0.028 & 0.055          \\
  (V1-1)-(V3-16) & -0.153  & 0.023 & 0.043          \\
  (V1-1)-(V2-8)  & -0.148  & 0.022 & 0.024          \\
  (V2-7)-(V3-18) & -0.155  & 0.024 & 0.043          \\
  (V2-9)-(V3-18) & -0.109  & 0.012 & 0.038          \\
  (V2-9)-(V3-15) & -0.042   & 0.002 & 0.036           \\
  (V1-1)-(V2-10) & -0.041   & 0.002 & 0.031           \\
  (V1-1)-(V3-15) & -0.001    & 0     & 0.069           \\
  (V1-2)-(V3-15) & -0.028   & 0.001 & 0.046           \\
  \hline
\hline
\end {tabular}
\end {table}
One can see that the hopping parameters between the occupied {\it xy} orbitals in the five-orbital
model are not exactly 
the same as in the one-orbital model. This is due to the fact that the hybridization between filled and 
vacant orbitals is explicitly taken into account in the five-orbital model.
It is also found that the sums of the squares of the hopping parameters between occupied and 
vacant orbitals I$^{f}_{ij}$ are larger than those between filled orbitals, I$^{a}_{ij}$.
Therefore one can expect that the Heisenberg model obtained from the one-band Hubbard 
model of Eq.(3) through perturbation theory is not sufficient  in order to describe the magnetic couplings 
between vanadium atoms in Na$_{2}$V$_{3}$O$_{7}$.

\section{LDA+U calculation}
The main goal of the LDA+U \cite{Anisimov} method is to take into account the Coulomb correlations 
of localized states, 
which are disregarded in LDA. The former method provides reasonable results for strongly correlated 
systems. \cite{korotin,Anisimov} 
The effective Coulomb interaction U and the effective intraatomic exchange J$_{H}$, which represent 
external parameters in a 
self-consistent cycle of the LDA+U scheme, are determined from the first-principle calculation 
by the constrained LDA method. 
This calculation scheme was described elsewhere. \cite{gunn}
The Coulomb interaction parameter U and intraatomic exchange J$_{H}$ have been estimated 
as U=2.3 eV and J$_{H}$=0.98 eV.
The Brillouin zone integration in our calculation has been performed in the grid 
generated by using (4,4,4) divisions.

\begin {table}[!h]
\centering
\caption [Bset] {Results of LDA+U calculations for different types of magnetic configurations (see Fig.4).}
\label {basisset}
\begin {tabular}{lcc}
  \hline   & conf. 1   & \quad \quad \quad conf. 2\\
  \hline
 Energy Gap (eV)                & 1.14  & \quad \quad \quad 1.18  \\
 Magnetic moment ($\mu_{B}$)    & 0.87  & \quad \quad \quad 0.89  \\
      \hline
\hline
\end {tabular}
\end {table} 

Our LDA+U calculations show that the magnetic structure 
of Na$_{2}$V$_{3}$O$_{7}$ is frustrated with respect to magnetic couplings, i.e. we cannot find a
collinear
spin configuration which corresponds to the minimum of the total energy of the system. 
We have performed the LDA+U calculations for two magnetic configurations (Fig.4). The results of these 
calculations are presented in Table III. One can see that in both magnetic 
configurations Na$_{2}$V$_{3}$O$_{7}$ 
is an insulator with an energy gap of 1 eV. The calculated average value of the magnetic moment 
of vanadium atoms is 
about 0.88 $\mu_{B}$, which is in good agreement with experimentally observed 
spin-$\frac{1}{2}$ moment.\cite{Gavilano} 
\begin{figure}[!h]
\includegraphics[width=0.4\textwidth]{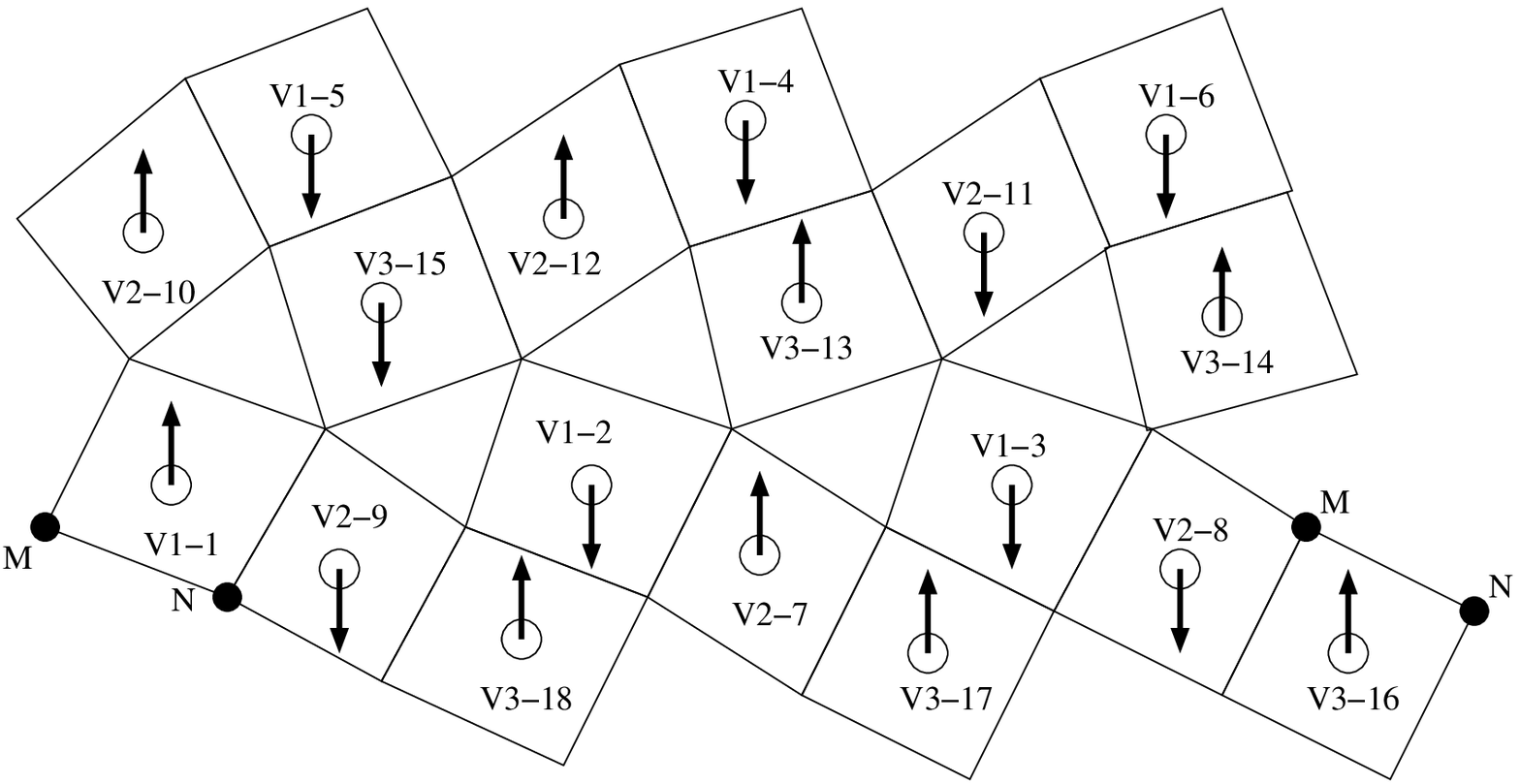}
\includegraphics[width=0.4\textwidth]{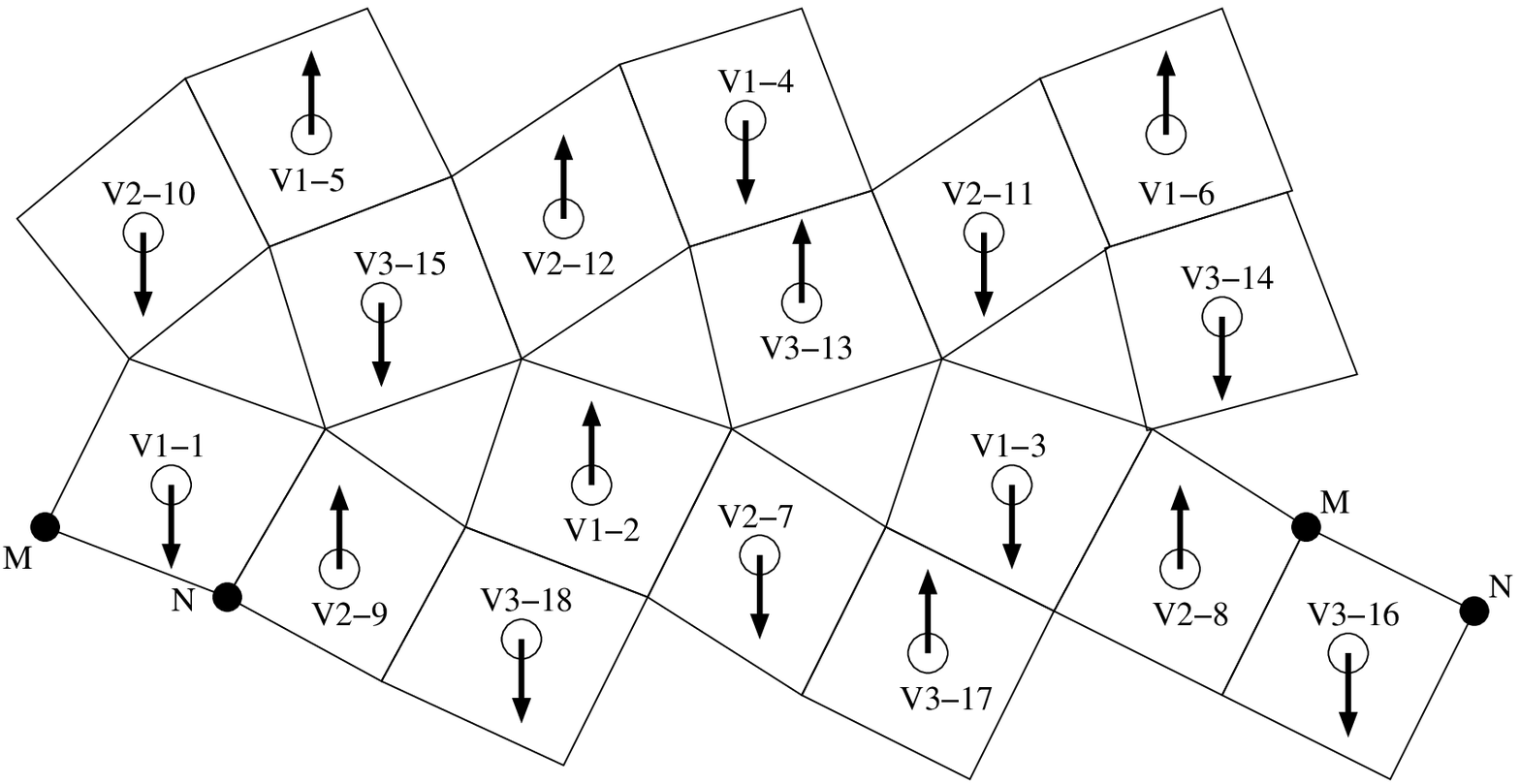}
\caption {The different magnetic configurations for Na$_{2}$V$_{3}$O$_{7}$ used in the LDA+U
calculation.
The arrows denote the directions of the magnetic moments, which lie along the $z$ axis.}
\label{total}
\end{figure}

Following Lichtenstein {\it et al.} \cite{Liechtenstein}, we determine the exchange interaction 
parameter between vanadium atoms
via the second variation of the total energy with respect to small deviations of the magnetic moments 
from the collinear magnetic configuration.
The exchange interaction parameters J$_{ij}$ of the spin 
Hamiltonian $H \, = \, \sum_{i > j} \, J_{ij} \, \vec S_{i} \vec S_{j}$, where S=$\frac{1}{2}$,  
can be written in the following form \cite{Liechtenstein,Mazurenko}:
\begin{eqnarray}
J_{ij} = \frac{2}{\pi} \int_{-\infty}^{E_{F}} d\epsilon \, {\rm Im} \sum_{\substack {m, m' \\ m'', m'''}} 
(\Delta^{mm'}_{i} \,
G_{ij \, \downarrow}^{m'm''} \, \Delta^{m'' m'''}_{j} \, G_{ji \, \uparrow}^{m''' m}) \nonumber
\end{eqnarray}
where the on-site potential $\Delta^{mm'}_{i}=H^{m m'}_{ii \, \uparrow} - H^{m m'}_{ii \, \downarrow}$ 
and  the Green function is calculated in the following way 
\begin{eqnarray}
G^{mm'}_{ij \sigma}(\epsilon) \, = \, \sum_{k,\, n} \frac{c^{mn \, *}_{i \sigma} \, (k) \, c^{m'n}_{j \sigma} \,
(k)}{\epsilon-E^{n}_{\sigma}} 
\end{eqnarray}
Here $c^{mn}_{i\sigma}$ is a component of the {\it n}th eigenstate, and 
E$_{\sigma}^{n}$ is the 
corresponding eigenvalue.
In order to define the contributions to the total exchange
interactions between vanadium atoms from different pairs of 3d orbitals, we diagonalize the
on-site potential in the following way:
$\Delta_{i}^{m \, m'}=A_{i}^{m \, k} \, \widetilde{\Delta}_{i}^{k \, k} \, (A_{i}^{k \, m'})^{*}$,
where $\widetilde{\Delta}_{i}^{kk}$ is the on-site potential which is diagonal in orbital space and 
the matrix A defines the basis where $\Delta$ is diagonal.
The exchange interaction between the {\it k}th orbital of site i and the {\it k$^{\prime}$}th orbital of 
site j is given by
\begin{eqnarray}
J^{kk'}_{ij} = \frac{2}{\pi} \, \int_{\infty}^{E_{F}} \, d\epsilon Im \, (\widetilde{\Delta}^{kk}_{i} \,
\widetilde{G}^{kk'}_{ij \, \downarrow} \, \widetilde{\Delta}^{k'k'}_{j} \, \widetilde{G}^{k'k}_{ji \, \uparrow}),
\nonumber
\end{eqnarray}
where $\widetilde{G}^{kk'}_{ij \, \sigma}= \sum_{m m'} A_{i}^{k \, m} \, 
G^{mm'}_{ij \, \sigma} \, (A_{j}^{m' \, k'})^{*}$.
In  Eq.(5), the index $n$ runs over all states. Therefore we take into account all interaction paths 
between the {\it i}th and {\it j}th atoms. 

Our calculated values of the total exchange couplings are presented in Table IV.
\begin {table}[!h]
\centering
\caption [Bset] {Calculated exchange interaction parameters of Na$_{2}$V$_{3}$O$_{7}$ (in meV). 
Negative couplings denote ferromagnetic exchange interactions. 
J$^{AF}_{ij}$ is the contribution to the total exchange interaction coming from the 
coupling between filled orbitals of the {\it i}th and {\it j}th atoms, while
J$^{FM}_{ij}$ is the sum of the exchange interactions between filled and vacant orbitals 
of the {\it i}th and {\it j}th sites.}
\label {basisset}
\begin {tabular}{lcccc}
  \hline
  i-j               & \quad \quad J$_{ij}$  & \quad \quad J$^{AF}_{ij}$ & \quad \quad J$^{FM}_{ij}$\\
  \hline
  (V1-1)-(V2-9)     & \quad \quad 31.8  & \quad \quad 54.2   & \quad \quad -22.4  \\
  (V1-1)-(V3-16)    & \quad \quad 13.4  & \quad \quad 35.4   & \quad \quad -22.0  \\
  (V1-1)-(V2-8)     & \quad \quad 22.3  & \quad \quad 34.5   & \quad \quad -12.2  \\
  (V2-7)-(V3-18)    & \quad \quad 15.5  & \quad \quad 35.6   & \quad \quad -20.1  \\
  (V2-9)-(V3-18)    & \quad \quad  -5.8   & \quad \quad 16.3   & \quad \quad -22.1  \\
  (V2-9)-(V3-15)    & \quad \quad  -12.2  & \quad \quad $\sim$ 0  & \quad \quad -12.2  \\
  (V1-1)-(V2-10)    & \quad \quad  -14.0    & \quad \quad $\sim$ 0  & \quad \quad -14.0  \\
  (V1-1)-(V3-15)    & \quad \quad -36.9   & \quad \quad 0.4    & \quad \quad -37.3  \\
  (V1-2)-(V3-15)    & \quad \quad -27.7   & \quad \quad 0.6    & \quad \quad -28.3  \\
  \hline
\hline
\end {tabular}
\end {table}
One can see that there is a qualitative {\it and} quantitative disagreement between the
exchange interactions J$^{xy}_{ij}$ deduced from the hopping parameters (Table I) and 
the total exchange interaction integrals J$_{ij}$ obtained by the Green 
function method (Table IV). 
Therefore the simple model estimation 
$J_{ij}=\frac{4 \, (t_{ij}^{xy})^{2}}{U}$, which was used in Ref.\onlinecite{saha-dasgupta}, does not work 
in the case of Na$_{2}$V$_{3}$O$_{7}$. 
One has to take into account the low symmetry of this compound. 
We have calculated the different contributions to the intra-ring and inter-ring exchange couplings.
It was found that the antiferromagnetic exchange interaction between occupied orbitals 
J$^{AF}_{ij}$ (Table IV) is in good agreement with $J^{xy}_{ij}$ deduced from the 
hopping integrals (Table I). 
But there are also strong ferromagnetic contributions which are
due to the hybridization between occupied (by electrons) and empty 3d orbitals of the vanadium atoms.
For vanadium atoms which belong to different rings, one can see that the antiferromagnetic 
contribution due to the hybridization 
between occupied {\it xy} orbitals is approximately zero (Table IV). 
Thus, the main contribution to the total exchange interaction is given by hybridization between occupied 
and empty orbitals.
A similar situation where the hybridization between filled and empty orbitals plays 
an important role was encountered 
for the layered vanadates CaV$_{2}$O$_{5}$, MgV$_{2}$O$_{5}$, CaV$_{3}$O$_{7}$, 
and CaV$_{4}$O$_{9}$\cite{Korotin99}. 

Let us first give an illustrative geometrical explanation of the obtained exchange interaction picture. 
In order to do so one can consider
the simple two pyramid S=$\frac{1}{2}$ model presented in Fig.5 with the assumption that the
$xy$ orbital of vanadium atom is occupied by one electron. 
\begin{figure}[!h]
\centering
\includegraphics[width=0.4\textwidth]{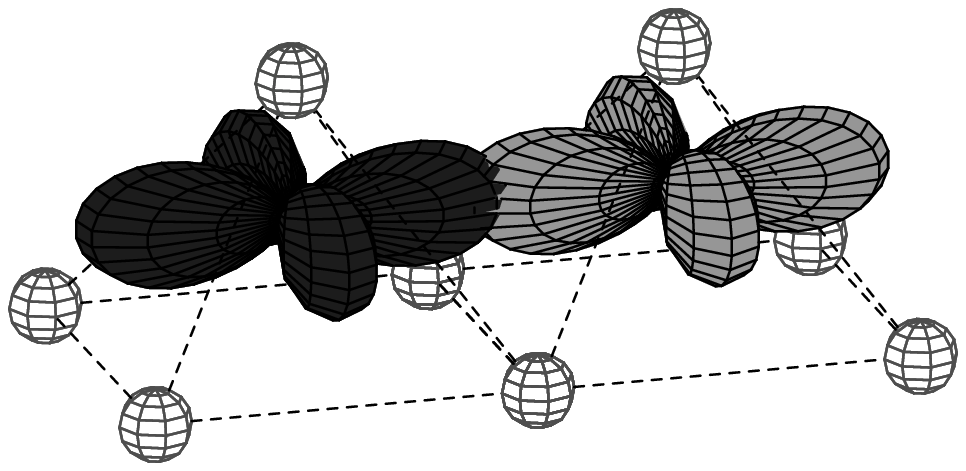}
\includegraphics[width=0.4\textwidth]{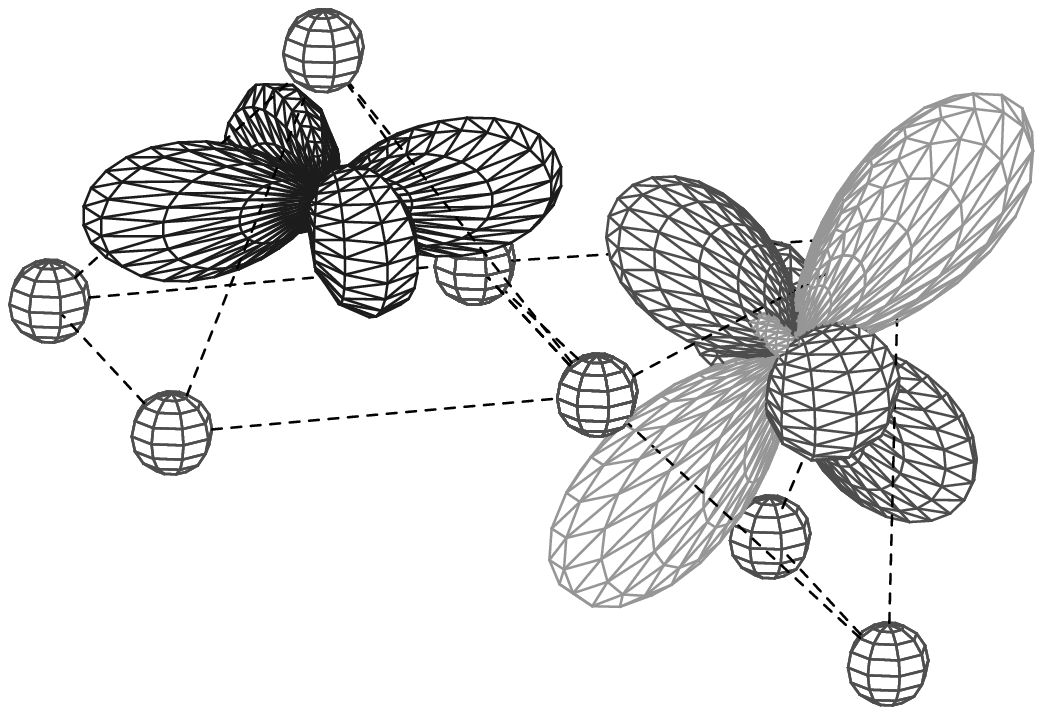}
\caption {Two pyramid {\it xy} model. The top and bottom figures correspond to perfect and 
distorted structures, respectively.
The oxygen atoms are located at the edges of pyramid.}
\end{figure}
The pyramid axes are collinear in the case of a perfect structure, therefore only the $\sigma$-overlap 
between the V $xy$-orbitals is non-zero. 
Thus the total exchange interaction is purely antiferromagnetic. 
The deviation from perfect geometry resulting from the rotation of a pyramid results in a
decrease of $\sigma$-overlap between 
active orbitals and an increase of those between occupied and vacant orbitals. For instance, 
the overlap between the vacant $3z^{2}-r^{2}$ (right pyramid) and the filled $xy$ orbital 
(left pyramid) results in an exchange interaction of ferromagnetic nature.

\begin {table}[!b]
\centering
\caption [Bset] {Calculated crystal-field splittings between the occupied ({\it xy}) and the vacant orbitals 
({\it yz}, {\it xz}, {\it 3z$^{2}$-r$^{2}$}, {\it x$^{2}$-y$^{2}$}) of vanadium atoms obtained by using 
the projection procedure (in eV).}
\label {basisset}
\begin {tabular}{lcccc}
  \hline
       & \quad \quad ${\bf \Delta}_{yz}$ & \quad \quad ${\bf \Delta}_{xz}$ & \quad \quad ${\bf \Delta}_{3z^{2}-r^{2}}$ & 
\quad \quad ${\bf \Delta}_{x^{2}-y^{2}}$   \\
  \hline
  V1         & \quad \quad 1.10         & \quad \quad  1.71        & \quad \quad 3.05            & \quad \quad 2.55     \\
  V2         & \quad \quad 1.39         & \quad \quad  1.08        & \quad \quad  2.96            & \quad \quad 2.41     \\
  V3         & \quad \quad 1.18         & \quad \quad  1.60        & \quad \quad 2.95            & \quad \quad 2.27     \\
  \hline
\hline
\end {tabular}
\end {table}

The resulting picture of magnetic interactions (Table IV) can be explained numerically within the
Kugel-Khomskii model.\cite{kugel} According to
their investigation, for real materials it is necessary to consider all the 3d levels and the intra-atomic 
exchange interaction J$_{H}$ which leads to Hund's rule. We modify the 
Kugel-Khomskii model by introducing 
the splitting between the occupied orbital (denoted as 1) and other vacant orbitals
\begin{eqnarray}
&H& = \sum_{\substack {i  j \sigma \\ m m' \\ (m \neq m')}}  \, 
t^{m m'}_{ij} a^{+}_{i m \sigma} a_{j m' \sigma} 
+ \sum_{\substack {i \sigma \sigma' \\ m m' }} 
\frac{U_{m m'}}{2}  n_{i m \sigma}  n_{i m' \sigma'} \nonumber \\
&\times& (1 - \delta_{m m'} \delta_{\sigma \sigma'})   
+ \sum_{i \, m' \, \sigma} (1-\delta_{1 \, m'}) \, 
(\varepsilon_{0} + {\bf \Delta}_{m'}) \, n_{i m' \sigma} \nonumber \\ 
&-& \sum_{\substack {i \sigma \sigma' \\ m m'}} 
\frac{J^{m m'}_{H}}{2} (1-\delta_{m m'}) \nonumber \\
&\times& (a^{+}_{i m \sigma} a_{i m \sigma'} a^{+}_{i m' \sigma'} a_{i m' \sigma} + 
a^{+}_{i m \sigma} a_{i m' \sigma} a^{+}_{i m \sigma'} a_{i m' \sigma'}),
\end{eqnarray}  
where $\varepsilon_{0}$ is the energy of the filled orbital and $(\varepsilon_{0} + {\bf \Delta}_{m'})$
is the energy of the $m'$th vacant orbital.  
For simplicity we assume that $\varepsilon_{0} = 0$ and we neglect the fact that different 
empty orbitals have different splittings, i.e. we assume that
${\bf \Delta}_{m'} = {\bf \Delta}$ (for vacant orbitals). It is also assumed that the Coulomb repulsion 
and the intraatomic exchange interation do not depend on the particular orbital, i.e. 
U$_{m m}$ = U$_{m m'}$ = U and J$^{m m'}_{H}$ = J$_{H}$.
The total exchange interaction can be expressed in terms of the 
sum of squares of the hopping integrals I$^{a}_{ij}$ and I$^{f}_{ij}$ presented in Table II, the splittings 
between filled and empty orbitals of vanadium atoms ${\bf \Delta}$,  
the on-site Coulomb U, and the intraatomic exchange interactions J$_{H}$ in the following form:
\begin{eqnarray}
J_{ij} = J^{AF}_{ij} + J^{FM}_{ij}  \quad \quad \quad \quad \quad 
\quad \quad \quad \quad \quad \quad \quad \nonumber \\
= \frac{4 I^{a}_{ij}}{U} - \frac{4 I^{f}_{ij} J_{H}}{(U+{\bf \Delta}) (U+{\bf \Delta}-J_{H})}. 
\end{eqnarray}  
The first term describes the antiferromagnetic coupling due to hybridization between active 
orbitals (occupied by electrons) at the {\it i}th and {\it j}th sites.
The second term is the sum of the exchange interactions between vacant and filled 
orbitals. It is of ferromagnetic nature.

\begin {table}[!b]
\centering
\caption [Bset] {Calculated contributions to the total exchange interactions between vanadium atoms 
obtained by using extended Kugel-Khomskii model (in meV).}
\label {basisset}
\begin {tabular}{lcc}
  \hline
  i-j  & \quad \quad \quad $J^{AF}_{ij}$   & \quad \quad \quad $J^{FM}_{ij}$   \\
  \hline
  (V1-1)-(V2-9)  & \quad \quad \quad 48.7  & \quad \quad \quad -26.2         \\
  (V1-1)-(V3-16) & \quad \quad \quad 40    & \quad \quad \quad  -20.5         \\
  (V1-1)-(V2-8)  & \quad \quad \quad 38.2  & \quad \quad \quad  -11.4        \\
  (V2-7)-(V3-18) & \quad \quad \quad 41.7  & \quad \quad \quad  -20.5        \\
  (V2-9)-(V3-18) & \quad \quad \quad 20.8  & \quad \quad \quad  -18.1         \\
  (V2-9)-(V3-15) & \quad \quad \quad 3.4   & \quad \quad \quad  -17.2       \\
  (V1-1)-(V2-10) & \quad \quad \quad 3.4   & \quad \quad \quad  -14.7         \\
  (V1-1)-(V3-15) & \quad \quad \quad  0    & \quad \quad \quad  -32.8         \\
  (V1-2)-(V3-15) & \quad \quad \quad 1.7   & \quad \quad \quad  -21.9         \\
  \hline
\hline
\end {tabular}
\end {table}

The values of the crystal-field splittings of different types of vanadium atoms 
obtained by using the projection procedure 
are presented in Table V.
If we take for the the splitting ${\bf \Delta}$ the average value of the crystal field splittings, namely 
2 eV, the ferromagnetic contributions  to the total exchange interactions 
estimated within the extended Kugel-Khomskii model are about 
2 times smaller than those obtained using the Green function method 
(presented in Table IV).  
A physically more appealing approximation is to use in the model calculation the smallest value of 
the crystal field splitting, which corresponds to the
first excitation energy for an electron between occupied and vacant orbitals. In our case this 
splitting is about 1.1 eV.      
The calculated contributions to the total exchange interactions between vanadium atoms for 
${\bf \Delta}$ = 1.1 eV are presented in Table VI.
One can see that we can qualitatively reproduce the anti- and ferromagnetic contributions to the
total exchange interactions between vanadium atoms
using this simple model except for bond (V2-9)-(V3-18). In order to obtain full qualitative agreement 
with the Green function method results one 
should use ${\bf \Delta}$ = 0.8 eV, which is slightly less than the crystal-field splitting calculated 
by the projection procedure.  

\section{CONCLUSION}
We have  investigated the  electronic structure and the microscopic origin of the exchange interactions
of Na$_{2}$V$_{3}$O$_{7}$. Based on our LDA+U results we predict that the value of the energy gap 
in electronic excitation spectra in this system is about 1 eV.
The different contributions to the  exchange interaction parameters have been calculated using the
 first-principle Green-function method
proposed by A.I. Lichtenstein. On the basis of the obtained results we argue that it is 
{\it not} correct to assume that the exchange interaction which is due to hybridization between active 
orbitals is a good approximation to the total exchange interaction between vanadium atoms 
for Na$_{2}$V$_{3}$O$_{7}$. 

According to the present results, the main interactions are antiferromagnetic along the ring, ferromagnetic
between the rings, and of the same order of magnitude. The resulting Heisenberg model is strongly 
frustrated, with no obvious starting point for perturbation theory, hence no simple way to predict
its properties. Such an analysis is left for future investigation.

\section{ACKNOWLEDGMENTS}
We would like to thank A.I. Lichtenstein, I.V. Solovyev,
and M.A. Korotin for helpful discussions.
The hospitality of the Institute of Theoretical Physics of EPFL is gratefully acknowledged.
This work is supported by INTAS Young Scientist Fellowship Program Ref. Nr 04-83-3230, 
Netherlands Organization for Scientific Research 
through NWO 047.016.005, the scientific program ``Development of scientific potential of universities'' section 3.3 project code 315, 
Russian Foundation for Basic Research grant RFFI 04-02-16096 and RFFI 03-02-39024.
The calculations were performed on the computer cluster of ``University Center of Parallel Computing'' of USTU-UPI. We also acknowledge the financial support of the Swiss National Fund and of MaNEP.

\end{document}